\documentclass[review]{elsarticle}

\usepackage{epsfig}
\usepackage{color}
\usepackage{mathtools}
\usepackage{hyperref}
\usepackage{ulem}
\usepackage[utf8]{inputenc}

\begin{document}

\begin{frontmatter}

\title{On the contact conditions for the density and charge profiles
in the theory of electrical double layer: from planar to spherical and
cylindrical geometry}

\author[ICMP]{Myroslav Holovko \corref{corresp}}
\cortext[corresp]{Corresponding author}
\ead{holovko@icmp.lviv.ua}

\author[FCCT]{Vojko Vlachy}

\author[ENSCP]{Dung di Caprio}

\address[ICMP]{Institute for Condensed Matter Physics,
National Academy of Sciences\\1 Svientsitskii Str., 79011 Lviv, Ukraine}

\address[FCCT]{Faculty of Chemistry and Chemical Technology, University of Ljubljana,\\
Vecna pot 113, SI-1000 Ljubljana, Slovenia}

\address[ENSCP]{Chimie ParisTech, PSL Research University, CNRS,
Institut de Recherche de Chimie Paris (IRCP), F-75005 Paris, France}

\begin{abstract}
In this paper, starting from the Bogoliubov-Born-Green-Yvon equations of the liquid-state
theory, we formulate two equivalent approaches for the calculation of the total density profile and of the
charge density profile of ionic fluids near nonplanar charged surfaces.
In the framework of these approaches, we establish exact conditions, that a
particular point of these profiles should satisfy, in the form of contact
theorems. These contact theorems for the total density profile and the charge density profile are obtained
by direct integration of a system of equations derived from the Bogoliubov-Born-Green-Yvon equations.
The contact theorems for both profiles have nonlocal character.
It is shown that the contact value of the total density profile
for uncharged
surfaces is characterized by the bulk pressure and the surface tension.
The contact theorems are applied to the cases of spherical and cylindrical surfaces.
It is shown that the contact theorem for the total density profile coincides
with the recent results obtained by W.~Silvester-Alcantara,
D.~Henderson and L.B.~Bhuiyan (\textit{Mol. Phys.}, 113, 3403, 2015)

\end{abstract}
\end{frontmatter}

The contact theorems are one of the few exact results in the theory of the electrical double layer.
They establish the exact expressions for the contact values of the total density profile (TDP) and of the charge density profile (CDP) at
the interface between an electrolyte and a charged electrode.
The contact theorem (CT) for the TDP in a planar electrical double layer
formed by a primitive model of electrolyte and a uniformly charged planar hard electrode was formulated
for the first time
more than fourty years ago by D.~Henderson, L.~Blum and J.L.~Lebowitz \cite{HendersonBlum,HendersonBlumLebowitz}
According to their results, the contact value of the total density
profile is given by the sum of the bulk osmotic pressure $P$ of the electrolyte and the
Maxwell stress tensor contribution 

\begin{eqnarray}\label{eq:CT1}
  \rho^{ct} &\equiv&  \sum_{\alpha}\rho_{\alpha} \left(\frac{d_\alpha}{2}\right) \nonumber\\
   &=& \beta P + \beta \frac{\varepsilon E^2}{8\pi}
\end{eqnarray}
where $\rho_{\alpha}(z)$ are the density distribution functions for ions of type
$\alpha$ at a normal distance $z$ from the wall,
$d_{\alpha}$ are the ion diameters,
$\beta=1/(k_BT)$ with $k_B$ the Boltzmann constant, $T$ the absolute temperature, $\varepsilon$
is the dielectric constant of the solvent{, where $E$ is the electric field} and $\displaystyle\frac{\varepsilon E}{4\pi}=q_s$
is the surface charge density per unit area on the wall.
Note that the same dielectric constant $\varepsilon$ is assumed in the electrode and in the electrolyte.
In these conditions, there are no image charges and therefore no image effects that need to be taken into account
between the electrode and the electrolyte. An example of an inhomogeneous dielectric constant will be discussed at the
end of the paper.

A little later in Ref.~\cite{CarnieChan}, a formal
derivation of the CT for the TDP was obtained
integrating directly the Bogoliubov-Born-Green-Yvon (BBGY)
\cite{Yvon,Bogoliubov,BornGreenBook} equation between the singlet and the pair
distribution functions for electrolytes in the presence of a charged electrode.
In our previous papers Refs.~\cite{MHJPBDDC,chargecontact2,chargecontact4},
in the framework of a similar approach based on the integration of the BBGY
equation, we formulated the CT for the CDP for the
planar electrical double layer.
The contact value of the CDP
can be presented in the following form
\begin{eqnarray}\label{eq:CTq1}
  q^{ct} &\equiv&  \sum_{\alpha} e_{\alpha}\rho_{\alpha} \left(\frac{d_\alpha}{2}\right)\nonumber\\
         &=&  \beta \sum_{\alpha} e_{\alpha}^2 \int_{d_\alpha/2}^\infty \rho_{\alpha}(z) \frac{\partial \psi (z)}{\partial z} dz
                    + \beta \sum_{\alpha} e_{\alpha} P_{\alpha}
\end{eqnarray}
where $e_{\alpha}$ is the charge of the ion of type $\alpha$,
$\psi (z)$ is the electrical potential defined as 
\begin{eqnarray}\label{eq:psi1}
   \psi (z) = - \frac{4\pi}{\varepsilon}\int_z^{\infty} q(z_1) (z_1-z) dz_1
\end{eqnarray}
where
\begin{eqnarray}\label{eq:q1}
   q (z) = \sum_{\alpha} e_{\alpha} \rho_{\alpha} (z)
\end{eqnarray}
is the CDP, $P_{\alpha}$ is the bulk partial pressure for the ions of type $\alpha$.
In contrast to the CT for the TDP,
the contact value of the CDP has a nonlocal character, the right hand side of Eq.~\ref{eq:CTq1}
involves an integral which requires the explicit knowledge of the ionic density profiles
accross the inhomogeneous region from the interface into the bulk.

For the symmetrical electrolyte with $d_+=d_-=d$ and $e_+=-e_-=e$, the second term
in Eq.~(\ref{eq:CTq1}) equals zero due to the electroneutrality condition.
As a result, the CT Eq.~(\ref{eq:CTq1}) reduces to the more simple form
\begin{eqnarray}\label{eq:CTq2}
q^{ct} =  \beta e^2 \int_{d/2}^\infty \rho(z) \frac{\partial \psi (z)}{\partial z} dz
\end{eqnarray}
where $\rho(z)$ is the TDP.

From the analysis of the effects of the ionic size, on the properties of the diffuse
double layer, using experimental and computer simulations results for
symmetrical electrolytes and for small charge at the wall, the semi
empirical local expression for the contact value $q^{ct}$, the
so called Fawcett-Henderson-Boda (FHB) conjecture \cite{FawcettHenderson,HendersonBoda} has been
proposed. According to this expression
\begin{eqnarray}\label{eq:CTqapprox1}
  q^{ct} =  \beta e^2 \frac{1}{\kappa} PE
\end{eqnarray}
where $\kappa$ is the inverse Debye length.

The connection between Eq.~(\ref{eq:CTq2}) and (\ref{eq:CTqapprox1})
was the subject of discussions in Refs.~\cite{BhuiyanHenderson,chargecontact5}.
In particular for a symmetrical electrolyte at small surface
charge, it was shown that both expressions give similar results which are in good agreement with computer
simulation data \cite{BhuiyanHenderson}.
In Ref.~\cite{chargecontact5}, it has been established that Eq.~(\ref{eq:CTqapprox1}) can be obtained from
Eq.~(\ref{eq:CTq2}) as a result of two approximations.
The first approximation is the result of the replacement in the integral in Eq.~(\ref{eq:CTq2})
of the TDP $\rho(z)$ by its contact value Eq.~(\ref{eq:CT1}).
As a result,
\begin{eqnarray}\label{eq:CTqapprox2}
  q^{ct} =  \beta e^2 \rho^{ct} \psi\left(\frac{d}{2} \right)
\end{eqnarray}
where $\psi(d/2)$ is the potential at the surface.
The second approximation is the consequence of the use of the linearized Gouy--Chapman theory \cite{Gouy,Chapman}
for $\psi(d/2)$.

The generalization of the CTs for charged nonplanar surfaces
is one of the challenges in the theory of the electrical double
layer.  In particular, after the formulation of the CT
for the TDP in the case of the planar double layer, Eq.~(\ref{eq:CT1}),
significant efforts {have been} focused on the generalization of this CT
for double layers with spherical, cylindrical and more complex nonplanar
geometries.  We note that according to the Henderson-Abraham-Barker approach
\cite{HendersonAbrahamBarker} the surface can be viewed as the surface of an
additional particle in the infinite dilution limit.
As a consequence, the TDP is equal to the pair distribution
function between this additional particle representing the wall and the ionic fluid.
It was also shown in Refs.~\cite{JRHenderson1983,JRHendersonJSRowlinson1984}
that in order to take into account the surface curvature effect, a surface term
should be added to the pressure in Eq.~(\ref{eq:CT1}).
This means that the excess chemical potential of the additional particle
can be written
\begin{eqnarray}\label{eq:CTcurv01}
  \mu_g^{ex} = P \mathcal{V}_g + {\gamma_g}\left(\frac{d}{2}+R_g\right) S_g
\end{eqnarray}
where $\gamma_g(r)$ is the surface tension defined as the surface part of the
chemical potential, $\mathcal{V}_g$ is the free volume accessible to the ions,
$S_g$ is the surface area of the curved wall.
In the following, the subscript $g$ will denote the geometry with $g=s$ or $c$
corresponding respectively to the spherical or cylindrical wall with the radius
$R_g=R_s$ or $R_g=R_c$ respectively.

We note that an expression similar to Eq.~(\ref{eq:CTcurv01}) was also used in the
scaled particle theory (SPT) \cite{ReissSPT1,ReissSPT2,HolovkoPatsahanDong2012}
for the description of the chemical potential of
the scaled particle in the hard sphere fluid.
In a similar way as in the SPT, ${\gamma_g}(d/2 + R_g)$ can be
expanded around $(d/2)/(d/2+R_g)$.
In the case of the hard sphere fluid near a spherical surface,
Bryk et al.  \cite{BrykRothMeckeDietrich2003} obtained the following expression
for the contact value of the TDP near an uncharged wall ($E=0$)
\begin{eqnarray}\label{eq:CTcurv02}
\hspace*{-0.60cm}  \rho_{s,E=0}^{ct} &=& \rho_s^{\mathit{HS}}\left(\frac{d}{2}+R_s\right)\nonumber\\
&=& \beta P^{\mathit{HS}} -\frac{9\eta^2}{4\pi(1-\eta)^3(d/2)^3}
   \left\{(1+\eta)\frac{d/2}{R_s+d/2} -\eta \frac{(d/2)^2}{(R_s+d/2)^2} \right\}
\end{eqnarray}
where $\rho_{s,E=0}^{ct}$ is the contact value of TDP
in the case of the spherical uncharged surface,
\textit{HS} superscript refers to the hard sphere system value of the corresponding quantity,
where $\eta=\frac{1}{6}\pi \rho d^3$ is the packing fraction of the hard sphere fluid.
As frequently done, in the SPT approach \cite{ReissSPT1,ReissSPT2,HolovkoPatsahanDong2012},
the hard sphere pressure is given by the isothermal compressibility
within the Percus-Yevick approximation.

Similar to Ref.~\cite{BrykRothMeckeDietrich2003}, the expression for the contact
value of the TDP for the hard sphere fluid near an uncharged cylindrical
wall can be derived and takes the following form
\begin{eqnarray}\label{eq:CTcurv03}
  \rho_{c,E=0}^{ct} &=& \rho_c^{\mathit{HS}}\left(\frac{d}{2}+R_c\right)\nonumber\\
  &=& \beta P^{\mathit{HS}} - \frac{9\eta^2}{4\pi(1-\eta)^3(d/2)^3}
   \left\{\frac{1}{2} (1+\eta) \frac{d/2}{R_c+d/2} \right\}
\end{eqnarray}
where is $\rho_{c,E=0}^{ct}$ is the contact value of the TDP
in the case of the cylindrical uncharged surface.

We should note that in the case of ionic fluids near an uncharged wall,
the pressure $P$ and the surface tension $\gamma_g(r)$ will include also
the corresponding coulombic contributions.

The second point, in order to generalize the CT
Eq.~(\ref{eq:CTcurv01}), is to calculate
the second term in Eq.~(\ref{eq:CTcurv01}) for the nonplanar cases.
So far, investigations have focused on the generalization of the
Maxwell tensor for the double layer with charged spherical and cylindrical
surfaces, charged membranes and charged surfaces with more complex geometries
\cite{VlachyBratko1981,Vlachy1982,Wennerstrom1982,Trizac1997,TellezTrizac2015}.
However, in these publications, the ionic sizes were not taken into
account neither explicitly nor implicitly in the calculation of the Maxwell
stress tensor.
It was shown that for electrolytes with point ions, the Maxwell stress tensor
contribution to the contact values of the TDP for walls with
different geometries has a similar form as in the case of the planar wall.

The exact formulation of the CT for the TDP for the
electric double layers in spherical and cylindrical geometries was obtained by
Silvestre-Alcantara, Henderson and Bhuiyan \cite{SilvestreHendersonBhuiyan2015}.
According to their results, the CT for the TDP for the
surface with nonplanar geometry can be presented in the form
\begin{eqnarray}\label{eq:CTcurv04}
  \rho_g^{ct} = \rho_{g,E=0}^{ct}
   -\beta E \int_{R_g+d/2}^{\infty} q(r)dr
\end{eqnarray}
where $q(r) = \sum_{\alpha} e_{\alpha}\rho_{\alpha}(r)$ is the CDP,
$\rho_{g}^{ct}$ is the contact value of the TDP for the geometry $g$ and
$\rho_{g,E=0}^{ct}$ is the corresponding contact value for the uncharged surface.
The hard sphere contribution to the
contact value of the TDP is given by Eq.~(\ref{eq:CTcurv02})
and (\ref{eq:CTcurv03}) for spherical and cylindrical surfaces
respectively.
As we can see the second term in Eq.~(\ref{eq:CTcurv04}) has a nonlocal character
and is more universal  than the first term.

In Ref.~\cite{SilvestreHendersonBhuiyan2015}, it was shown that in the limit
$R_g \rightarrow \infty$, the CT Eq.~(\ref{eq:CTcurv04}) reduces to
\begin{eqnarray}\label{eq:CTcurv05}
  \rho^{ct} = \beta P + \frac{\beta E^2 \varepsilon}{4\pi}
\end{eqnarray}
which differs from Eq.~(\ref{eq:CT1}) by the second term which
has a missing factor $1/2$ compared to Eq.~(\ref{eq:CT1}).
The authors explained this difference by the fact that in the planar case,
in order to satisfy electroneutrality,
one must assume a second parallel electrode of opposite charge at infinite
separation from the first electrode.\\

In this note, we will show that the CT for the TDP
in the form Eq.~(\ref{eq:CTcurv04}) can be obtained from the BBGY equations.
Starting from the BBGY equations, we will also obtain the exact relation for the
CDP. For convenience, we present the BBGY relations in the following form
\begin{eqnarray}\label{eq:CTcurv06}
  \nabla_1\rho_\alpha(1) = -\beta\nabla_1 V_\alpha(1)\rho_\alpha(1)
   -\beta\rho_\alpha(1)\sum_\gamma \int \rho_\gamma(2)
    g_{\alpha\gamma}(12)\nabla_1 u_{\alpha\gamma}(r_{12})d2
\end{eqnarray}
where $1$ and $2$ stand for the position of the corresponding particles,
$V_\alpha(1) = e_\alpha V_\alpha^{coul}(1) + V_\alpha^{sh}(1) $ is
the external potential,
$u_{\alpha\gamma}(r_{12}) = e_\alpha e_\gamma u^{coul}(r_{12}) + u_{\alpha\gamma}^{sh}(r_{12})$
is the interaction pair potential.
The subscripts $\alpha$ and $\gamma$ stand for ions having charges
$e_\alpha$ and $e_\gamma$, the superscripts $coul$ and $sh$ indicate
the electrostatic and non electrostatic short-range contributions.
$\rho_\alpha(1)$ is the density distribution of the particle of type $\alpha$
at point $1$, $g_{\alpha\gamma}(12)$ is the pair distribution function
between particles of type $\alpha$ and $\gamma$ at positions 1 and 2.

We consider here two equivalent approaches, where in both cases, we start from
the BBGY equation.
In the first approach, we start from the BBGY equation in the form
\begin{eqnarray}\label{eq:CTcurv07}
  \nabla_1\rho_\alpha(1) = -\beta e_\alpha\rho_\alpha(1)\nabla_1 V^{coul}(1)
   -\beta\rho_\alpha(1)\nabla_1 V_\alpha^{sh}(1)
   -\beta\rho_\alpha(1)\nabla_1 w_\alpha(1),
\end{eqnarray}
where
\begin{eqnarray}\label{eq:CTcurv08}
  \nabla_1 w_\alpha(1) = \sum_\gamma \int d2 \rho_\gamma(2)
    g_{\alpha\gamma}(12)
\left[e_\alpha e_\gamma\nabla_1 u^{coul}(r_{12}) 
             + \nabla_1 u_{\alpha\gamma}^{sh}(r_{12})\right]
\end{eqnarray}
corresponds to the average force on a particle of type $\alpha$
at point $1$.

In this approach Eqs.~(\ref{eq:CTcurv07})-(\ref{eq:CTcurv08})
include explicitly the coulombic interactions between the ions and the surface,
$V^{coul}(1)$ and the direct interionic Coulomb interactions $u^{coul}(r_{12})$.
Due to the long range nature of the Coulomb interaction, both terms
taken separetely can give divergent terms.
Due to this we consider the second approach in a similar way
as in Refs.~\cite{CarnieChan,MHJPBDDC}, we introduce the mean electrostatic
potential $\psi(1)$ defined by
\begin{eqnarray}\label{eq:CTcurv09}
  \nabla_1 \psi(1) = \nabla_1 V^{coul}(1) +
\int \sum_\gamma e_\gamma \rho_\gamma(2)
  \nabla_1 u^{coul}(12) d2.
\end{eqnarray}

In this case, as in \cite{MHJPBDDC}, we can rewrite the BBGY equation in the
following form
\begin{eqnarray}\label{eq:CTcurv10}
  \nabla_1 \rho_\alpha(1) = -\beta e_\alpha \rho_\alpha(1) \nabla_1 \psi(1)
   - \beta \rho_\alpha(1) \nabla_1 V_\alpha^{sh}(1)
   - \beta \rho_\alpha(1) \nabla_1 W_\alpha(1)
\end{eqnarray}
where
\begin{eqnarray}\label{eq:CTcurv11}
  \nabla_1 W_\alpha(1) = \sum_\gamma \int d2 \rho_\gamma(2)
\left[e_\alpha e_\gamma h_{\alpha\gamma}(12)\nabla_1 u^{coul}(r_{12})       
             + g_{\alpha\gamma}(12) \nabla_1 u_{\alpha\gamma}^{sh}(r_{12})\right]
\end{eqnarray}
corresponds to the average force on a particle of type $\alpha$ at point $1$
with the exception of the contribution from the mean electrical potential and
$h_{\alpha\gamma}(12)=g_{\alpha\gamma}(12) -1 $ is the 
pair correlation function.

From Eq.~(\ref{eq:CTcurv08}) and Eq.~(\ref{eq:CTcurv11}), we can see that
$w_{\alpha}(1)$ and $W_{\alpha}(1)$ are related by following equation
\begin{eqnarray}\label{eq:CTcurv11b}
\nabla_1 w_\alpha(1) = \nabla_1 W_\alpha(1) + \int \sum_\gamma e_\alpha e_\gamma \rho_\gamma (2) \nabla_1 u^{coul}(12)d2. 
\end{eqnarray}
Equations (\ref{eq:CTcurv07}) and (\ref{eq:CTcurv10}) are two
different equivalent versions of the BBGY equations.
In Eq.~(\ref{eq:CTcurv07}), the first term is the force on the ion
of charge $e_\alpha$ in a direction normal to the wall due to the
presence of the charged surface.
Due to the long range nature of the coulombic interaction
$V^{coul}(1)$, this force does not vanish in the bulk and leads to an
ill-defined divergent term in the equations.
However this term is cancelled by the second term in Eq.~(\ref{eq:CTcurv11b})
which also gives a similar coulombic contribution.
We do not have such issues if we start from Eq.~(\ref{eq:CTcurv10})
where the electrostatic force
$\nabla_1 \psi(1)$ defined by Eq.~(\ref{eq:CTcurv09}) includes at once both
coulombic contributions.
Due to this, it is better to use Eq.~(\ref{eq:CTcurv10}) than Eq.~(\ref{eq:CTcurv07}).

Now, we formulate the equations for the TDP and the CDP
\begin{eqnarray}\label{eq:CTcurv12}
  \rho(1) = \sum_\alpha \rho_\alpha(1)
\;\;\;\;\;\;\mbox{and}\;\;\;\;\;\;
  q(1) = \sum_\alpha e_\alpha \rho_\alpha(1).
\end{eqnarray}

In order to proceed, we can use the BBGY equation in the standard form Eq.~(\ref{eq:CTcurv07})
which includes the Coulomb interaction between wall and ions,
$V^{coul}(1)$, or in the form Eq.~(\ref{eq:CTcurv10}) which includes the mean electrostatic
potential $\psi(1)$. We note that in both cases, we can omit the term related
to the external potential $V_\alpha^{sh}(1)$ which we consider as equivalent
to that of a hard wall.
As a result from Eq.~(\ref{eq:CTcurv07}), we can formulate the following
system of equations for the TDP and the CDP
\begin{eqnarray}\label{eq:CTcurv13}
  \nabla_1\rho_\alpha(1) &=& -\beta\nabla_1V^{coul}(1)q(1)
    -\beta\sum_\alpha \rho_\alpha(1) \nabla_1 w_\alpha(1)\\
  \nabla_1 q(1) &=& -\beta\nabla_1V^{coul}(1)\hat{\rho}(1)
    -\beta\sum_\alpha e_\alpha \rho_\alpha(1) \nabla_1 w_\alpha(1)
         \label{eq:CTcurv14}
\end{eqnarray}
where 
\begin{eqnarray}\label{eq:CTcurv15}
   \hat{\rho}(1) = \sum_\alpha e_\alpha^2\rho_\alpha(1).
\end{eqnarray}

After integrating Eq.~(\ref{eq:CTcurv13}) and Eq.~(\ref{eq:CTcurv14}) in space, from the surface
into the bulk,
we obtain the CT for the TDP and for the CDP.
These expressions generalize
the corresponding relations we obtained in Ref.~\cite{MHJPBDDC} for the planar case.
We note that in the first term in
Eq.~(\ref{eq:CTcurv13}) and in Eq.~(\ref{eq:CTcurv14}), the Coulomb potential
reads $V^{coul}_s(1)\propto\displaystyle\frac{1}{\varepsilon r} Q_s$ for spherical surfaces,
and $V^{coul}_c(1)\propto\displaystyle-\frac{1}{\varepsilon} Q_c \ln(r/r_c)$ for cylindrical surfaces,
where $Q_s$ and $Q_c$ are the charges of the corresponding surfaces.
Since $\nabla_1 V^{coul}_s(1)\propto\displaystyle-\frac{1}{\varepsilon r^2} Q_s$
and $\nabla_1 V^{coul}_c(1)\propto\displaystyle-\frac{1}{\varepsilon r} Q_c$,
after integrating Eq.~(\ref{eq:CTcurv13}) and Eq.~(\ref{eq:CTcurv14}),
we obtain the generalization of the CTs for the TDP and the
CDP near a nonplanar wall. Hereafter, for simplicity, we assume $d_+=d_-=d$
and we obtain the CTs in the following forms
\begin{eqnarray}\label{eq:CTcurv16}
\rho_g^{ct} &=& \rho_{g,E=0}^{ct}
        -\beta E \int_{R_g+d/2}^\infty q_g(r) dr   \\
   q_g^{ct} &=&
      \sum_\alpha e_\alpha \rho_{g,E=0,\alpha}^{ct}
      -\beta E\sum_\alpha\int_{R_g+d/2}^\infty \hat{\rho}_\alpha(r) dr
                \label{eq:CTcurv17}
\end{eqnarray}
where $\rho_g^{ct}$ and $q_g^{ct}$ are the contact values of the total
density and of the charge density for surfaces of type $g$ and
where $\rho_{g,E=0,\alpha}^{ct}$ is the contact value of the density profile for
the corresponding uncharged surface and for particle of type~$\alpha$.

We note that the CT Eq.~(\ref{eq:CTcurv16}) for the density
profile is identical to Eq.~(\ref{eq:CTcurv04}) obtained by
Silvester-Alcantara, Henderson and Bhuiyan \cite{SilvestreHendersonBhuiyan2015}.

Starting from the BBGY equation in the form~(\ref{eq:CTcurv10}), which
includes the mean electrostatic potential $\psi(1)$ instead of the coulombic
potential $V^{coul}(1)$, we can write Eq.~(\ref{eq:CTcurv13})
and  Eq.~(\ref{eq:CTcurv14}) for the TDP and the CDP in the following
equivalent forms
\begin{eqnarray}\label{eq:CTcurv18}
  \nabla_1\rho_\alpha(1) &=& -\beta q(1)\nabla_1 \psi(1)
    -\beta\sum_\alpha \rho_\alpha(1) \nabla_1 W_\alpha(1)\\
  \nabla_1 q(1) &=& -\beta\hat{\rho}(1) \nabla_1 \psi(1)
    -\beta\sum_\alpha e_\alpha \rho_\alpha(1) \nabla_1 W_\alpha(1).
         \label{eq:CTcurv19}
\end{eqnarray}

Now after integration of these equations, we can present
the CTs for the TDP and for the CDP near
a nonplanar wall in the following form
\begin{eqnarray}\label{eq:CTcurv20}
   \rho_{g}^{ct} &=& \rho_{g,E=0}^{ct}
        -\beta \int_{R_g+d/2}^\infty q_g(r) \frac{\partial \psi(r)}{\partial r} r^{\delta_g}dr \\
   q_{g}^{ct} &=&
\sum_\alpha e_\alpha \rho_{g,E=0,\alpha}^{ct}
   -\beta \int_{R_g+d/2}^\infty \hat{\rho}(r) \frac{\partial \psi(r)}{\partial r}  r^{\delta_g} dr
                \label{eq:CTcurv21}
\end{eqnarray}
where $\delta_g=2$ for spherical wall, $\delta_g=1$ for cylindrical
wall and $\delta_g=0$ for planar wall, where we have included in the list of
geometries also the case $g=p$ for the planar wall.

In a similar way as for Eqs.~(\ref{eq:CTcurv07}) and
(\ref{eq:CTcurv10}) the ensuing Eqs.~(\ref{eq:CTcurv16})-(\ref{eq:CTcurv17}) and
(\ref{eq:CTcurv20})-(\ref{eq:CTcurv21}) have different structures.
The equivalence of the two different forms (\ref{eq:CTcurv16})
and (\ref{eq:CTcurv20}) of the CT for the TDP for the planar double
layer has been proved by Henderson and Blum \cite{HendersonBlum2}.
This proof can be generalized to the nonplanar case. Due to symmetry
properties, $\nabla_1 u^{coul}(12)=-\nabla_2 u^{coul}(12)$ and the electroneutrality condition,
at least for symmetrical ionic systems, reads
\begin{eqnarray}\label{eq:CTcurv21b}
- \beta \sum_\alpha \int e_\alpha \rho_\alpha (1) \int \sum_\gamma e_\gamma \rho_\gamma (2) \nabla_1 u^{coul}(12)\;d1d2 =0
\end{eqnarray}
which proves the equivalence of the forms (\ref{eq:CTcurv16}) and (\ref{eq:CTcurv20})
of the CT for the TDP.

Concerning the CT for the charge, owing to the divergence of the
second term of the CT for the CDP in
Eq.~(\ref{eq:CTcurv17}), due to the integrals $\int_{R_g+d/2}^\infty \hat{\rho}_\alpha(r) dr$,
the CT for the CDP in the form (\ref{eq:CTcurv17})
cannot be used and the term
\begin{eqnarray}\label{eq:CTcurv21c}
- \beta \sum_\alpha \int e_\alpha^2 \rho_\alpha (1) \int \sum_\gamma e_\gamma \rho_\gamma (2) \nabla_1 u^{coul}(12)\;d1d2
\end{eqnarray}
in equation (\ref{eq:CTcurv17}) should be taken into account which results in the second form
of the CT Eq.~(\ref{eq:CTcurv21}).

For charge symmetric electrolytes $e_+=-e_-=e$, the {first} term in
the CT for the CDP  in the form~(\ref{eq:CTcurv17})
or (\ref{eq:CTcurv21}) vanishes.
Moreover $\hat{\rho}(1)$ is proportional to the TDP 
\begin{eqnarray}\label{eq:CTcurv22}
   \hat{\rho}(1) = e^2 \rho(1)
\end{eqnarray}
and we have a simple nonlocal relation between the contact value of the
CDP and the TDP
\begin{eqnarray}\label{eq:CTcurv24}
   q_{g}^{ct} &=&  - \beta e^2 \int_{R_g+d/2}^\infty \rho(r)
    \frac{\partial \psi}{\partial r}\,r^{\delta_g}\,dr
\end{eqnarray}
which generalizes Eq.~(\ref{eq:CTq2}) for the nonplanar cases.

In the limit $R_s\rightarrow\infty$ for the spherical surface or in the limit
$R_c\rightarrow\infty$ for the cylindrical surface, the CTs for
the TDP and for the CDP reduce to the planar case.
In particular in \cite{SilvestreHendersonBhuiyan2015},  it was shown that due
to the local electroneutrality condition
\begin{eqnarray}\label{eq:CTcurv25}
  -\int_{d/2}^\infty q(z) dz = \frac{\varepsilon E}{4\pi},
\end{eqnarray}
the last term in Eq.~(\ref{eq:CTcurv16}) for the contact value
of the TDP can be presented, in this limit, in the following form
\begin{eqnarray}\label{eq:CTcurv26}
  -\beta E \int_{R_g+d/2}^\infty  q(r) dr \xrightarrow[R_g\rightarrow\infty]{} -\beta E \int_{d/2}^\infty q(z) dz = \beta\frac{\varepsilon E^2}{4\pi}
\end{eqnarray}
which differs from the corresponding term in Eq.~(\ref{eq:CT1}) by a factor $1/2$.

In Ref.~\cite{SilvestreHendersonBhuiyan2015}, this result was explained  by the fact
that in the case of a planar symmetry, one must assume the existence of a second
electrode of opposite charge at infinite separation from the first electrode which
contributes the identical additional electric field $\varepsilon E/(8 \pi)$. In order to
obtain the similar result from Eq.~(\ref{eq:CTcurv20}) we should assume that 
\begin{eqnarray}
 - \int^\infty_{R_g+d/2} q_g(r)\frac{\partial \psi(r)}{\partial r} r^{\delta g} dr
   \xrightarrow[R_g\rightarrow\infty]{}
     -2 \beta \int_{d/2}^\infty q(z) \frac{\partial \psi(z)}{\partial z} dz
   = \beta \frac{\varepsilon E^2}{4\pi}.
\end{eqnarray}
In an analogue way to Eq.~(\ref{eq:CTcurv24}), in this limit, we can then
reproduce the result of Eq.~(\ref{eq:CTq2}).
In Ref.~\cite{SilvestreHendersonBhuiyan2015}, it was noted that the electroneutrality
condition for the spherical and the cylindrical double layers has respectively
the following forms
\begin{eqnarray}\label{eq:CTcurv26b}
  - \int_{R_s+d/2}^\infty  q_s(r) r^2 dr = \frac{R_s^2\varepsilon E^2}{4\pi},\\
  - \int_{R_c+d/2}^\infty  q_c(r) r dr = \frac{R_c\varepsilon E^2}{4\pi}.
\end{eqnarray}
Due to this, for the nonplanar case, the CT for the TDP
cannot be presented in simple local forms as for the planar case.

We should note that, the last term in the CT in the form~(\ref{eq:CTcurv20}),
due to the Poisson equation
\begin{eqnarray}\label{eq:CTcurv27}
   \nabla_{1}^{2}  \psi(1) = -\frac{4\pi}{\varepsilon}q(1)
\end{eqnarray}
can be presented also in the form
\begin{eqnarray}\label{eq:CTcurv28}
    - \beta \int_{R_g+d/2}^\infty q_g(r) \frac{\partial \psi}{\partial r} r^{\delta_g}dr
  = \beta \frac{\varepsilon}{4\pi}
\int_{R_g+d/2}^\infty \nabla_{1}^{2}  \psi_g(1)  \nabla_{1} \psi_g(1) r^{\delta_g}dr
\end{eqnarray}
where $\psi_g$ is the mean force potential for surface of type $g$ (=$c$ or $s$).

Since
\begin{eqnarray}\label{eq:CTcurv29}
   \nabla^{2}  \psi_g(1) = \frac{\partial^2 \psi_g(1)}{\partial r^2 }
    + \frac{\delta_g}{r} \frac{\partial \psi_g(1)}{\partial r}
\end{eqnarray}
we see that the integral in Eq.~(\ref{eq:CTcurv28}) cannot give the simple
Maxwell stress term as in the case of the planar wall \cite{MHJPBDDC}.
For the planar case $\delta_g=0$ and in Eq.~(40) after integration it is easy to
obtain the Maxwell term of Eq.~(1).\\

In this paper, starting from the BBGY equations between the singlet and pair
distribution functions of ionic fluids near a charged planar surface,
we formulate two approaches in order to calculate the charge and the
total density ionic profiles.
The starting point are the two equivalent equations~(\ref{eq:CTcurv07}) and (\ref{eq:CTcurv10}).
The first equation includes the direct Coulomb interaction between the ions and the surface.
The long-range character of this interaction leads to a force on a fixed ion 
which is normal to the wall and  which does not vanish in the bulk. However, this term is
cancelled by the force created by the long-range coulombic interactions between the
fixed ion and the surrounding ions. As a result, the partial Maxwell electrostatic
stress tensor includes both the contributions from the ion-wall interaction and from
the interionic interactions.
In the second approach, using the Poisson equation, the direct ion-surface interaction is transformed
into the mean electrical potential.
The CTs for the TDP and the CDP are then formulated
for planar and nonplanar surfaces by direct integration of Eq.~(\ref{eq:CTcurv07}) and Eq.~(\ref{eq:CTcurv10}).
The equations which are obtained are then applied to the cases of spherical and cylindrical surfaces.
For the uncharged surfaces, the contact value of the TDP is found to be
characterized by the bulk pressure and the surface tension.

In both approaches, the TDP contact values described by equations
(\ref{eq:CTcurv16}) and (\ref{eq:CTcurv20}) are equivalent and identical to
the result recently formulated by Silvestre-Alcantara, Henderson and Bhuiyan
\cite{SilvestreHendersonBhuiyan2015}.
However the first approach leads to an ill-defined expression for
Eq.~(\ref{eq:CTcurv17}) due to the divergence of the CT for
the CDP.
In the second approach, this divergence is cancelled by contributions from
interionic Coulomb interactions.
As a consequence, only the equation obtained in the second approach,
Eq.~(\ref{eq:CTcurv21}), can be used.\\
In the present paper, we have obtained exact expressions for the 
TDP and for the CDP for the planar, spherical and cylindrical surfaces. Most
expressions obtained for the CTs have nonlocal character.
Only the CT for the TDP in the planar double layer is local.
This means that obtaining a local expression is rather an exception.
We note that in the case of the planar double layer, the CT for the
CDP can be reduced to the local form in the framework of the
FHB conjecture \cite{FawcettHenderson,HendersonBoda}.
When compared to simulation results, this conjecture proves quite effective at low surface charge densities.
The FHB conjecture can also be applied to the CT for the CDP in
the case of a nonplanar double layer.
As a result, the CT  Eq.~(\ref{eq:CTcurv24}) for the CDP
for charge symmetrical electrolytes reads
\begin{eqnarray}
   q_{g}^{ct}= - \beta e^2 \rho_{g}^{ct} \int_{R_g+d/2}^\infty \frac{\partial \psi(r)}{\partial r} r^{\delta g} dr
\end{eqnarray}
where $\rho_{g}^{ct}$ has a nonlocal form and is given by one of the
equivalent forms Eq.~(\ref{eq:CTcurv16}) or Eq.~(\ref{eq:CTcurv20}).
The expressions obtained for the CTs are complex and have a hybrid character,
they include both local and nonlocal terms.
For example, the expressions for the contact value of the TDP
includes three contributions, namely the bulk fluid pressure, the surface
tension and the electrostatic terms. The first two contributions are local.
The surface tension term is strongly dependent on the surface geometry and has a
different form for spherical and cylindrical surfaces. The electrostatic term is
in general nonlocal and presented in two different
equivalent forms Eq.~(\ref{eq:CTcurv16}) or Eq.~(\ref{eq:CTcurv20}).
In Eq.~(\ref{eq:CTcurv16}), the electrostatic term is defined by the volume
independent integral which has the same form for the planar, cylindrical and
spherical geometries.
In the second form, the electrostatic term is presented by
the volume dependent integral which has different forms for different
geometries.
For the contact value of the CDP, only Eq.~(\ref{eq:CTcurv21})
obtained in the second approach is well-defined
as opposed to Eq.~(\ref{eq:CTcurv17})
derived in the first approach which has a diverging term and is ill-defined.
For the charge symmetrical case, this equation
only includes the electrostatic term and takes the form Eq.~(\ref{eq:CTcurv24}).
The electrostatic term has a form similar as that for the contact
value of the total density profile and is defined by the volume dependent integral
which has a different form for different surface geometries.

We also note that for the simple symmetrical electrolytes, the TDP and the CDP
can be presented in the following forms
$\rho(r)=\rho g_s(r)$, $q(r)=\rho e g_d(r)$,
where
$g_s(r)=(1/2)(g_+(r)+g_-(r))$, $g_d(r)=(1/2)(g_+(r)-g_-(r))$, and where
$g_+(r)$ and $g_-(r)$ are the singlet distribution functions for counterions and coions respectively.
Using the CTs Eq.~(\ref{eq:CTcurv20}) and  Eq.~(\ref{eq:CTcurv24}),
we can obtain the following expressions for the contact values of the singlet
distributions of counterions and coions respectively
\begin{eqnarray}\label{eq:CTcurv50}
\rho g_+^{ct}=\frac{1}{2} \rho_{g,E=0}^{ct}- \beta e \int_{R_g+d/2}^\infty \rho g_+(r) \frac{\partial \psi(r)}{\partial r} r^{\delta g} dr
\end{eqnarray}
\begin{eqnarray}\label{eq:CTcurv51}
\rho g_-^{ct}=\frac{1}{2} \rho_{g,E=0}^{ct} + \beta e \int_{R_g+d/2}^\infty \rho g_-(r) \frac{\partial \psi(r)}{\partial r} r^{\delta g} dr.
\end{eqnarray}
These expressions generalize our previous results obtained in Ref.~\cite{MHJPBDDC} for the planar surface.
Accordingly, the expressions for the contact values of the profiles
for counterions and coions near charged surfaces are given by the bulk fluid
pressure, the surface tension and an integral which includes the product of the
electrostatic potential and the profile of counterions and coions respectively.
The expressions obtained Eq.~(\ref{eq:CTcurv50}) and Eq.~(\ref{eq:CTcurv51}) are
exact and can be used to improve the integral equation description of the
electrolyte in the spherical and the cylindrical double layer
\cite{Lozada1,Lozada2,Lozada3} and also to improve the
description of the double layer properties in the framework of the density
functional approach \cite{Gillespie1,Pizio1,Pizio2}.
We believe that, in both approaches, it is possible to obtain a better evaluation
of thermodynamic and electrophysical properties such as the differential
capacitance, for example.

Finally, we note that the existence of a nonplanar geometry is not the
only reason for the nonlocal formulation of the CT for
the TDP.
Another reason can be connected with the nonhomogenity of the interionic interaction
near the surface. A typical example of such inhomogenity appears when the dielectric
constant of the surface $\varepsilon_s$ is significantly different from the dielectric
constant of the solvent $\varepsilon$.
Hereafter, the subscript $s$ stands for surface.
We need to solve the Poisson equation for the electric potential $\psi(r)$ and for
the electrical displacement taking into account the discontinuity of the
dielectric at the boundary $r=R_g+d/2$.  The electric displacement is continuous
at the boundary and verifies
$\displaystyle\left.\varepsilon  \frac{\partial \psi(r)}{\partial r} \right|= \left. \varepsilon_s \frac{\partial \psi(r)}{\partial r} \right|_s$.
This equation can be solved using the image charge method
\cite{jackson1999classical} which leads to extra terms for the
ion-wall interaction $V_{\alpha,im}^{coul}(1)$ and for the interionic interactions $u_{\alpha\gamma,im}^{coul}(1,2)$.
This is a well known result for a single planar dielectric discontinuity
\cite{jackson1999classical,yukhnovsky1980statistical,Golovko1985}.
The interactions are now given by
\begin{eqnarray}
V_{\alpha,im}^{coul}(1)= \left(\frac{\varepsilon_s - \varepsilon }{\varepsilon_s + \varepsilon}\right) \;\frac{e_\alpha^2}{2 \varepsilon (z+d/2)}
\end{eqnarray}
and
\begin{eqnarray}
u_{\alpha\gamma,im}^{coul} (1,2)= e_\alpha e_\gamma u_{im}^{coul} (r_{12,im})
\end{eqnarray}
with
\begin{eqnarray}
u_{im}^{coul}(r_{12,im})=- \left(\frac{\varepsilon_s - \varepsilon }{\varepsilon_s + \varepsilon}\right) \frac{1}{\varepsilon r_{12,im}},
\end{eqnarray}
where $r_{12,im}$ is the distance between the point $r_1$ and the image charge of the second ion at point $r_2$.
In the case of a slit pore, the two dielectric discontinuity create multiple image
effects. The real charges create images on either sides of the slit, and in turn these
images generate subsequent images as a consequence alternatively of one or the other interface.
The resulting expressions for the interactions have the form of an infinite sum of contributions
\cite{torrie1982electrical}.
Similar problems for spherical and cylindrical geometries were considered in
\cite{messina2002image,cui2006electrostatic} respectively. The inclusion of the
images effects for the CT for the density leads to two nonlocal
supplementary terms in the Eq.~(\ref{eq:CTcurv16}) or
Eq.~(\ref{eq:CTcurv20})
\begin{eqnarray}
&-& \beta \int d1 \sum_\alpha \rho_\alpha(1)\nabla_1 V_{\alpha ,im}(1) \nonumber\\
&&- \beta \int d1 \int d2 \sum_{\alpha,\gamma} e_\alpha e_\gamma \rho_\alpha(1)\rho_\gamma(2) h_{\alpha\gamma}(1,2)\nabla_1 u_{im}(r_{12,im})
\end{eqnarray}
and generalizes the result obtained by Carnie and Chan \cite{CarnieChan} for the
planar case to a nonplanar geometry.\\

In the present paper we have focused on the formulation of the CTs for
the TDP and for the CDP
in the interface between an electrolyte and a
charged hard nonplanar electrode. In the future, we plan to generalize
the results we have obtained to a liquid electrode.
In this case, we need to pay attention due to the
soft nature of the surface between the electrode and the electrolyte.
By using our previous result for the CTs for anisotropic
fluids near a hard planar wall \cite{contacttheoremaniso},
we also intend to the extend our previous results to the case of
orientationally ordered charged dispersions.

\section*{Acknowledgments}
M. Holovko gratefully acknowledges financial support from the National Research Foundation of Ukraine
(project N 2020.02/0317).
V. Vlachy acknowledges the support of the  Slovenian Research Agency (ARRS) funding through project J1-1708.

\renewcommand{\baselinestretch}{1.55} \small \normalsize
\biboptions{numbers,sort&compress}
\bibliography{biblio}

\end{document}